\documentclass[useAMS,usenatbib]{mn2e}
\usepackage{graphicx} 
\usepackage{pslatex}
\usepackage{bm}
\usepackage[table]{xcolor} 
\usepackage{soul} 

\title[Weak lensing and the Hubble constant]
{The influence of weak lensing on measurements of the Hubble constant with quad-image gravitational lenses}

\author[M. Jaroszynski and J. Skowron]
{M. Jaroszynski$^{1}$\thanks{E-mail:
mj@astrouw.edu.pl (MJ); jskowron@astrouw.edu.pl (JS)} 
and J. Skowron$^{1}$\\
$^{1}$University of Warsaw Observatory, Al. Ujazdowskie 4, 00-478 Warsaw,
Poland}

\begin{document}

\date{Accepted; Received ;}


\definecolor{HighlightColor}{HTML}{00FFCF}
\sethlcolor{HighlightColor}
\newcommand{\modi}[1]{\hl{#1}}

\maketitle

\label{firstpage}

\begin{abstract}
We investigate the influence of matter along the line of sight 
and in the strong lens vicinity on the properties of 
quad image configurations and on the measurements of the 
Hubble constant ($H_0$). We use simulations of light propagation in 
a nonuniform universe model with the distribution of matter in space 
based on the data from Millennium Simulation. For 
a given strong lens and haloes in its environment we model the matter 
distribution along the line of sight many times, using different 
combinations of precomputed deflection maps representing subsequent 
layers of matter on the path of rays. We fit the simulated quad image 
configurations with time delays using nonsingular isothermal ellipsoids 
(NSIE) with external shear as lens models, treating the Hubble constant 
as a free parameter. 
We get a large artificial catalog of lenses with derived values 
of the Hubble constant, $H^\mathrm{fit}$.  The average and median 
of  $H^\mathrm{fit}$ differ from the true value used in
simulations by $\le 0.5~\mathrm{km/s/Mpc}$ which includes 
the influence of matter along the line of sight  and in the lens
vicinity, and uncertainty in lens parameters, except the slope of the
matter distribution, which is fixed.
The characteristic uncertainty of $H^\mathrm{fit}$ is $\sim
3~\mathrm{km/s/Mpc}$. Substituting the lens shear parameters with values
estimated from the simulations reduces the uncertainty to $\sim
2~\mathrm{km/s/Mpc}$. 

\end{abstract}


\section{Introduction}

The measurement of the Hubble constant based on the cosmic ladder
has a long history (see e.g. \citealp{freedmad10}, \citealp{riess11} for
reviews). 
The methods based on CMB anisotropy (\citealt{WMAP9}, \citealt{planckXVI}) 
give a high formal accuracy of $H_0$ derivation 
but are not in full agreement with each other.
On the other hand measurements based on the gravitational lenses 
with time delays (\citealp{refsdal64}) have their own attraction, at least 
as a consistency check, since they are independent of other methods.

In the idealized case of an isolated lens with a known mass distribution 
profile placed in (otherwise) uniform universe (\citealp{refsdal64}) 
the accuracy of the Hubble constant derivation depends only on the 
accuracy of the time delay measurement and is straightforward. 
In a more realistic approach 
one has to take into account other mass concentrations, since strong 
gravitational lenses are typically observed in complex environments 
(see e.g. \citealp{Will06}).

\citet{oguri07} gives the value of $H_0$ ($68\pm6\pm8~\mathrm{km/s/Mpc}$) 
based on 16 lensed QSOs. The values for individual systems have a large 
spread, but the claimed statistical and systematical errors 
for the sample are of the order of 10\%. Similarly \citet{paraficz10} 
find $H_0$ based on 18 systems to be $68^{+6}_{-4}~\mathrm{km/s/Mpc}$ 
but $76^{+3}_{-3}~\mathrm{km/s/Mpc}$ when they use a sub-sample of 5
elliptical lenses with an extra constraint on their mass profiles, which
illustrates the importance of systematical errors involved.
\citet{rathna15} obtain $68\pm6~\mathrm{km/s/Mpc}$ based on 10 systems.
The systematic errors are not estimated.

\citet{Suy10} obtain Hubble constant ($H_0=70.6\pm3.1~\mathrm{km/s/Mpc}$)
with a single, well constrained observationally strong lens B1698+656
using a cosmological model with fixed density parameters.
The matter distribution in the lens environment and along the line of
sight is modeled in details based on abundant observations. 
\citet{Suy13} use two lenses and WMAP data to constrain cosmological
parameters. The Hubble constant is found with 4\% -- 6\% accuracy
depending on the assumptions on cosmological model.
Similarly \citet{fadely10} employ the first gravitational lens
Q0957+561, to obtain $H_0=79.3^{+6.7}_{-8.5}~\mathrm{km/s/Mpc}$.

The influence of the mass distribution in the lens vicinity and along
the line of sight on its models has been investigated by many authors 
(\citealp{KA88}, \citealp{KKS97}, \citealp{barkana96}, \citealp{ChKK03},
\citealp{WBO04}, \citealp{WBO05}, \citealp{Mom06}, \citealp{Aug07}
and \citealp{AN11} to cite a few). \citet{KZ04} investigate the influence 
of the lens environment on derived model parameters (including $H_0$) 
using a synthetic group of galaxies. 
The problem of the environment influence is investigated in their
work by \citet{Suy10}, \citet{Wong11}, \citet{Suy13}, \citet{Collett13},
\citet{McCully14}, \citet{McCully16} among others. A mixture of
observational, numerical, and statistical methods is used to improve the
accuracy of the external shear and convergence estimates.

The description of light travel in a nonuniform universe model, which 
addresses part of the problems arising in connection with $H_0$ derivation,
is still being improved. 
\citet{McCully14}, \citet{McCully16}
 using a multiplane approach, assume that in 
most layers the deflection can be modeled as shear, but they allow for 
more than a single layer, where the full treatment can be applied. This
approach saves computation time by using shear approximation in majority of 
layers but its algebra is rather complicated. 
\citet{dAloisio13} develop a formalism based on \citet{barkana96} approach,
effectively improving the {\it single lens plus external shear model} 
at the cost of introducing some additional parameters. \citet{Sch14a},
\citet{Sch14b} discusses the mass sheet degeneracy in the multiplane context
and investigates the accuracy of cosmological parameters derived from
its modeling.

In this paper we continue our investigation of the environmental and line
of sight effects which influence the action of strong gravitational lenses.
Our calculations are based on the results of the Millennium Simulation 
\citep{Spr05} provided by its 
online database \citep{ls06}. The main purpose of this study is the 
evaluation of the influence of such effects on the accuracy of the 
measurement of the Hubble constant. The matter density distribution
obtained from the Millennium Simulation (or any other simulation
investigating gravitational instability on scales of a few hundred Mpc) is
not sufficient as a basis for strong lensing study by galaxies 
because of too low resolution \citep{hilbert07}. \citet{hilbert09} 
use the matter distribution from the Millennium Simulation to investigate 
weak lensing effects, but their methods are not directly applicable to 
our purposes. We follow an approach in many aspects similar to work of
\citet{Collett13}. We use the information on the distribution in space of
gravitationally bound haloes provided by \citet{b11} and \citet{b4} and
based on the Millennium Simulation. The
haloes are characterized by their virial masses, radii, and velocities
only. The mass distribution inside haloes, their ellipticity and
orientation in space have to be specified (see Sec.~\ref{raydefl}).
We use different density profiles for haloes as compared 
to \citet{Collett13}.

We follow the approach of \citet{JK12} (hereafter Paper I)
and \citet{JK14} (hereafter Paper II), changing our simulations methodology. 
More emphasis is put on the modeling of the matter 
distribution along the line of sight (hereafter LOS), which is key to assess
the systematic uncertainties in the Hubble constant measurement. We use the fact that 
matter distribution in space is uncorrelated on distances of hundreds of 
Mpc and model LOS as a random combination of many uncorrelated weak 
lenses between the source and the observer. Using a large number of such
combinations we get a large number of  LOS models, 
and its influence can be statistically investigated. With galaxies being 
neighbours of strong lenses (called environment, hereafter ENV) the problem 
is  more difficult, since they are  correlated with each other. 
We consider only one ENV model for each strong lens (based on the 
distribution of galaxies in its vicinity in the Millennium Simulation). 
We do not follow \citet{KZ04}, who get different environments switching 
the roles of the main lens and its neighbours.
Thus the number of  different ENV models is equal to the number of strong 
lenses considered (1920 in the reported investigation). Some measure of  ENV 
effects can also be obtained by comparing the results of simulations
including both, LOS and ENV, with the results based on the inclusion 
of LOS only. We also use the model with the main lens in a uniform universe
model (hereafter UNI) for comparison.

Our investigation concentrates on the influence of the matter along
the line of sight and the strong lens environment on the fitted values
of the Hubble constant. The mass profile of the lens (another major source
of errors in $H_0$ modeling) is not investigated here. 
The lenses are chosen at random from a set of sufficiently massive
Millennium haloes. On average their environments are poor as compared
e.g. to a set of six quad lenses investigated by \citet{Wong11}. This
suggests that the results of our approach maybe applicable to the 
samples of less extreme lenses which may be found  by ongoing large 
sky surveys.

In Sec.~\ref{sec:model} we describe our approaches to light propagation. 
Sec.~\ref{sec:quad}  presents tools used to compare different models and
the results of such comparison. Sec.~\ref{sec:fits} is devoted to the main problem of measuring Hubble constant 
based on several lenses with measured time delays. Discussion and conclusions 
follow in Sec.~\ref{sec:concl}.

\section[]{Model of the light propagation}
\label{sec:model}

\subsection{Ray deflections and time delays}
\label{raydefl}
We follow the methods of Papers I and II, using 
the multiplane approach to gravitational lensing (e.g. \citealp{SW88};
\citealp{SS92}) employing the results of the Millennium Simulation 
\citep{Spr05} and the non-singular isothermal ellipsoids (NSIE) as models 
for individual  haloes (\citealp{KSB94}; \citealp{K06}). 

In our approach the matter distribution is described as a {\it background} 
component represented by  matter density given on a low resolution $256^3$ 
grid plus gravitationally bound haloes given by \citet{b11} and \citet{b4}. 
The Millennium Simulation uses periodic boundary conditions, so
calculation of the gravitational acceleration based on known matter density
distribution and Fourier transform is straightforward in 3D. The
component of the acceleration perpendicular to the rays (with GR
correction factor) can be used to calculate the deflections and time
delays due to the background.

We use nonsingular isothermal ellipsoids (NSIE) to model matter
distribution in all haloes. The NSIE model as described 
by \citet{KSB94} gives the deflection and lens potential in analytical
form, but corresponds to infinite mass. In 2D real notation one has
(\citealp{K06}): 
\begin{eqnarray}
\alpha_x(x,y,\alpha_0,q,r_0)&=&
\frac{\alpha_0}{q^\prime}~\mathrm{arctan}
\left(\frac{q^\prime~x}{\omega+r_0}\right)\\
\alpha_y(x,y,\alpha_0,q,r_0)&=&
\frac{\alpha_0}{q^\prime}~\mathrm{artanh}
\left(\frac{q^\prime~y}{\omega+q^2r_0}\right)~\mathrm{, where}\\
\omega(x,y,q,r_0)&=&\sqrt{q^2(x^2+r_0^2)+y^2}~~~~q^\prime=\sqrt{1-q^2}
\end{eqnarray}
The ray crosses the lens plane at $(x,y)$, the lens center is placed at
the  origin of the coordinate system, the major axis along $x$. 
The axis ratio is given by $q$, $r_0$ is the core radius, and $\alpha_0$
is  the approximate asymptotic value of the deflection angle.

Each projected halo is represented as a difference between two
NSIE distributions with the same characteristic deflection angles $\alpha_0$,
axis ratios $q$, and position angles,
but different values of core radii $r_1 \ll r_2$, which makes its mass
finite: 
\begin{eqnarray}
\bm{\alpha}&=&
\bm{\alpha}(x,y,\alpha_0,q,r_1)-\bm{\alpha}(x,y,\alpha_0,q,r_2)\\
\lim_{r\rightarrow\infty}\bm{\alpha}&=&
\alpha_0(r_2-r_1)\frac{\bm{r}}{r^2}~~~~~\Leftrightarrow\\
M&=&\frac{c^2}{4G}\alpha_0(r_2-r_1)
\label{virialmass}
\end{eqnarray}
(compare Paper I). The above formula gives the value of characteristic 
deflection $\alpha_0$ for a halo of given mass and virial radius 
$r_\mathrm{vir}\approx r_2$. (We use $r_1 =0.001 r_2$ which guarantees
the smoothness of formulae at $r=0$ and has little impact on the whole lens.)
The axis ratios $q$ and position angles are not
given by cosmological simulations. For $q$ we assume a distribution
probability within the range $0.5 \le q \le 1$ with a maximum at $q=0.7$, 
loosely resembling  the results of \citet{KY07}. The position angles in 
the sky are random. 
Since the {\it background } contains the whole mass, including mass of
the haloes,
the latter must be "void corrected" by some negative density distribution. 
We use discs with the negative density approaching zero at the outer radius: 
\begin{equation}
\Sigma(r)=
-\frac{3M}{\pi r_\mathrm{d}^2}\left(1-\frac{r}{r_\mathrm{d}}\right)
\end{equation}
where $\Sigma$ is the surface mass density, $M$ is the mass of the halo and 
$r_\mathrm{d}$ - the radius of the negative density disc defined by 
$M=4/3\pi\rho r_\mathrm{d}^3$, where $\rho$ is the mean density in the
Universe at the epoch of interest.
(In Papers I and II we were using constant surface density discs, but 
the present approach avoids discontinuities at the edge).
A void corrected halo does not deflect rays outside its $r_\mathrm{d}$ 
radius, so only a finite number of haloes has to be included in calculations.

The {\it snapshots} of the Millennium Simulation, giving the matter
distribution in space, correspond to 
a discrete set of redshifts $\{z_i\}$. We follow this arrangement
placing our deflection planes at the same redshifts. Each deflection
plane represents the influence of matter in a layer perpendicular to the
rays with the thickness given by the distance traveled by photons
between consecutive planes.
For each layer we construct a {\it deflection map} giving two components of 
the deflection angle on a grid covering the region of interest.  Similarly 
we construct a {\it time delay map} representing the influence of 
gravitational potential of the layer.

The light travels a few hundred $\mathrm{Mpc}$ through each layer, so 
we do not expect any correlations between the distributions of matter 
belonging to different layers along a ray. Since before cutting out the layer,
we randomly shift and 
rotate Millennium cubes belonging to different epochs 
(\citealp{carbone08}), to avoid the consequences of periodic boundary 
conditions, such 
correlations are excluded in our approach, anyway. Thus choosing
a random path of rays through space means choosing one deflection map
and time delay map for every layer. Since the choice of 
locations of ray paths in different layers is independent, we can
apply the {\it prismatic transformation} \citep{gfs88} in every layer
without loosing generality. (The deflection in one layer influences 
the positions of a ray in all subsequent layers. As long as matter 
distributions in different layers are not correlated, this fact has no 
consequences.) We transform our maps in such a way
that the deflection angle at the middle point of any map vanishes:
\begin{equation}
\bm{\alpha}_i(\bm{\beta}_i) =
\bm{\alpha}_i^\prime(\bm{\beta}_i) -\bm{\alpha}_{i0}^\prime
~~~~~\Delta t(\bm{\beta}_i)=
\Delta t^\prime(\bm{\beta}_i)-d_i\bm{\alpha}_{i0}^\prime\cdot\bm{\beta}_i
\end{equation}
where the variables before transformation are denoted with primes, and 
$\bm{\alpha}_{i0}^\prime$ is the original deflection at the central point.
The subscript $i$ enumerates the layers, $\bm{\beta}_i$ gives the position 
in the $i$-th layer, and $d_i$ is the comoving distance to the layer. 
After the transformation the central ray (going through middle 
points of all maps) is not deflected at all and may be used as an axis
in a rectangular coordinate system. Propagation of light beams 
corresponding to different sets of deflection maps can be compared
when using such coordinate system.

\subsection{Deflection maps}

We use light beams of the 
$6^{\prime\prime}\times 6^{\prime\prime}$ crossection at the observer's
position. The deflection maps cover a slightly larger solid angle of
$10^{\prime\prime}\times 10^{\prime\prime}$ to allow for the beam 
deformations. The beams are wide enough to accommodate a typical image 
configuration resulting from a galaxy scale strong lens.

We use two kinds of maps for weak and strong lensing separately. Weak 
lensing maps represent deflections and time delays for a ray bundle 
traveling at random direction and starting at random location in 
a given layer. By chance the deflection on a map constructed in such a way
may not be weak, reflecting the possibility of finding another strong lens
along the line of sight. This has an impact on the results, but we do not 
reject such maps a'priori.

Strong lensing maps represent the deflections and time delays by a strong 
lens and its neighbours. The lenses with measured time delays
investigated by \cite{oguri07}, \cite{paraficz10}, and \cite{rathna15}
have redshifts range $0.26 \le z_\mathrm{L} \le 0.89$. The corresponding
sources belong to the redshift range $0.65 \le z_\mathrm{S}\le 3.60$.
We place our 
lenses on ten adjacent Millennium layers with a similar redshift range 
($0.32 - 0.83$).  Each strong lens is found
by looking for an appropriate halo close to a randomly chosen point 
inside the simulation cube, which, treated
as singular isothermal sphere, would give the Einstein radius between
$0.5^{\prime\prime}$ and $1.5^{\prime\prime}$ for a source at 
$z_\mathrm{S} \approx 2$. 
Finally we use randomly directed beam of rays
passing through and map deflections and time delays caused by the halo 
and its environment. On a separate map we store deflections and time 
delays due to the halo alone. 

We construct 16 strong lensing maps for each of the ten Millennium layers 
at $0.32\le z_i \le 0.83$, which is the assumed range of the lens 
redshifts. Similarly we calculate a sample of weak lensing 
maps covering the redshifts  $0 \le z_i \le 2.62$, which corresponds to
the possible range between the observer and the source.
There are 64 weak lensing  maps in every layer.
All maps use grids of the size $512\times 512$. The choice of the 
size and number of maps results from the memory capacity considerations; 
we are able to store such an atlas of strong and weak lensing maps in the RAM.
In order to increase the number of simulated cases, we repeat the whole 
process 12 times, every time creating a new atlas of independently 
calculated maps. Thus there are $16\times 10\times 12=1920$ strong
lensing maps each representing a different halo with its surroundings.
In principle there 
is no problem in using maps belonging to different atlases,
but it would be technically less efficient.

\subsection{Simulations of light propagation}

The multiplane approach describes the path of a ray as (eg
\citealt{SW88}): 
\begin{equation}
\bm{\beta}_N = \bm{\beta}_1 
- \sum_{i=1}^{N-1}~\frac{d_{iN}}{d_{N}}~\bm{\alpha}_i(\bm{\beta}_i)
\label{multiplane}
\end{equation}
where $\bm{\beta}_N$ is the position of the ray in the $N$-th layer,
$d_{ij}$ is the angular diameter distance as measured by an observer
at epoch $i$ to the source at epoch $j$. W also use the subscripts $O$, $L$,
and $S$ for the observer, lens, and source planes respectively,
and $d_i\equiv d_{Oi}$.  
$\bm{\alpha}_i(\bm{\beta}_i)$ is the deflection angle in the 
$i$-th layer at the position $\bm{\beta}_i$.
Since we consider sources at different redshifts we do not use {\it
scaled} deflection angles in the above and further equations as is
customary in the multiplane approach \citep{SEF92}.  
In a flat cosmological model
the angular diameter distances in the lens equation can be replaced by 
comoving distances, which we shall denote $D_{ij}$ with the same
subscript convention.
In the calculations we apply more efficient recurrent formula of
\citet{SS92}, equivalent to the above equation.

Knowing the light path, one can calculate the {\it geometric} part of the 
relative time delay $\Delta t_\mathrm{geom}$ (as compared with the 
propagation time along a null geodesics in a uniform universe model) using 
the formula \citep{SEF92}:
\begin{equation}
c\Delta t_N^\mathrm{geom}(\bm{\beta}_1) = 
\frac{1}{2}~\sum_{i=1}^{N-1}(1+z_i)~\frac{d_{i+1}}{d_id_{i,i+1}}
\left(d_i(\bm{\beta}_{i+1}-\bm{\beta}_i)\right)^2
\label{geomdelay}
\end{equation}
where we consider a ray coming to the observer from the direction 
$\bm{\beta}_1$ (which defines its earlier path, so all $\bm{\beta}_i$ are 
known) and the factors $1+z_i$ represent time dilatation.

The deflection in each layer can be calculated as a gradient of the deflection 
potential, which is also a measure of {\it gravitational} time delay 
$\Delta t_\mathrm{grav}$ \citep{SEF92}:
\begin{equation}
\bm{\alpha}_i = -\frac{1}{d_i}\frac{\partial\Psi_i}{\partial\bm{\beta}_i}
~~~~~~c\Delta t_\mathrm{grav}(\bm{\beta}_i) = \Psi_i(\bm{\beta}_i) + C 
\end{equation}
The potential is defined up to a constant $C$. The cumulative 
{\it gravitational} time delay after crossing all the layers is:
\begin{equation}
c\Delta t_N^\mathrm{grav}(\bm{\beta}_1)= 
\sum_{i=1}^{N-1}~(1+z_i)\Psi_i(\bm{\beta}_i)
\end{equation}
where again $\bm{\beta}_1$ defines the path and we account for time 
dilatation.
Finally:
\begin{equation}
\Delta t_N(\bm{\beta}_1) 
= \Delta t_N^\mathrm{geom}(\bm{\beta}_1) 
+\Delta t_N^\mathrm{grav}(\bm{\beta}_1) 
\end{equation}
The above expression contains an unknown additive constant. Only the 
difference in calculated time delays between two rays can have a clear 
physical meaning.

After choosing a set of the deflection maps (see below) we model the
backward propagation of a bundle of rays using Eq.~\ref{multiplane} for
each ray. The result of the ray shooting is a vector array:
\begin{equation}
\bm{\beta}_N^{kl}=\bm{\beta}_N(\bm{\beta}_1^{kl})
\end{equation}
where $\bm{\beta}_N^{kl}$ gives the positions in the source plane 
of rays apparently coming from the directions $\bm{\beta}_1^{kl}$ 
on the observer's sky. Superscripts $k$, $l$ enumerate the rays. 
Similarly the time delays are stored in an array $\Delta t_N^{kl}$.

Our simulations proceed as follows. 
We use all our strong lenses from the ten redshift layers within 
$0.32\le z_i\le 0.83$ range, a total of $16 \times 10 \times 12=1920$ 
maps representing main lenses with their surroundings which cause 
the ENV effects. For each lens we  draw a source redshift belonging to
$1.23 \le z_\mathrm{S} \le 2.62$, which covers most of the source
redshifts range mentioned above. Next we choose a random combination of 
weak lensing maps in layers between the observer and the main lens and between
the main lens and the source, which represent LOS effects. After that we
backward shoot a beam of $512\times 512$ rays covering the 
$6^{\prime\prime}\times 6^{\prime\prime}$ solid angle.
We check that all rays remain within the maps for all layers. We store the 
positions of rays in the source plane.  This is the type of simulation, 
dubbed LOS+ENV, which is the most realistic. We repeat every such calculation 
64 times, using different combinations of weak lensing maps, but keeping 
the strong lens, its environment, and source redshift unchanged.

For comparison we perform also very similar simulations using the same 
choice of weak lensing maps, but replacing the main lens with surroundings
by the same lens isolated. Such an approach is dubbed LOS. For each lens we
use the same 64 weak lens combinations as in LOS+ENV approach.

In the ENV simulations we use main lenses with surroundings, but remove 
weak lensing maps, thus neglecting the LOS effect. The lenses are physically 
correlated with their neighbours, so there is no room for making random 
combinations with other surroundings, and the number of cases
considered is much smaller as compared to approaches including LOS.
Finally using an 
isolated lens in a uniform universe model (UNI) we get another case for 
comparison. 

The methods described above investigate the results of strong
lensing with perturbations from LOS and/or ENV. 
For the interpretation of the results an independent estimate of
perturbing effects is needed. 
To find the external shear and convergence acting on ray bundles 
we use the weak lensing approximation. For each of our weak
lensing maps we calculate the derivative of the deflection angle 
$\bm{\alpha}$ with the respect the ray position $\bm{\beta}$ using 
finite differences on a scale of  $\sim 1~\mathrm{arcsec}$ similar to a
typical separation between strong lens images:
\begin{equation}
\mathsf{\Gamma}_i^\prime\equiv 
\left|\left|\frac{\Delta\bm{\alpha}_i}{\Delta\bm{\beta}_i}\right|\right|
\equiv \left|\left|\begin{array}{cc}
\kappa^\prime+\gamma_1^\prime & \gamma_2^\prime  \\
\gamma_2^\prime         & \kappa^\prime-\gamma_1^\prime
\end{array} \right|\right|
\label{weak_one}
\end{equation}
where $\kappa^\prime$, $\gamma_1^\prime$, and $\gamma_2^\prime$ give the
convergence and shear components defined up to a scaling factor
depending on the observer - lens - source distances.
Using the same method we calculate also weak lensing effect of the
strong lens environment, applying the above formula to the deflections
caused by the neighbours, but neglecting the main lens itself.

\begin{figure}
\includegraphics[width=84mm]{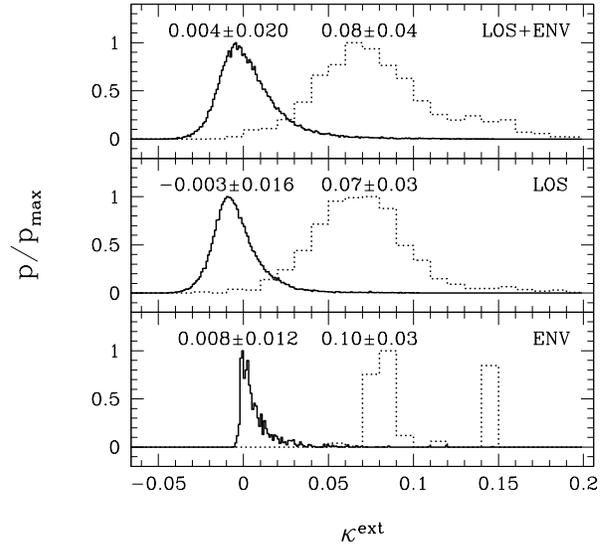}
\caption{Probability distribution of the external convergence caused 
by LOS and/or  ENV calculated in the weak lensing approximation 
(Eq.~\ref{weak_all}) for all cases resulting in successful fits of image
configurations (solid lines). For comparison with \citet{Suy13} Fig.~6 
the distributions weighted by the likelihood of 
$\gamma^\mathrm{ext}=0.089\pm0.006$ are shown with dotted lines. 
The average values and standard deviations are given above the
respective plots. 
}
\label{kappa}
\end{figure}

In the case of quad lenses, the source is close to the optical axis.
Using comoving distances in a flat Universe one can approximate a ray
belonging to a quad image as:
\begin{eqnarray}
\bm{\beta}_i&=& \bm{\beta}_1~~~~~~~~~~~~~~~
\left|\left|\frac{\partial\bm{\beta}_i}{\partial\bm{\beta}_1}\right|\right|
=\mathsf{I}
~~~~~~~~~~~~i\le n_{L}\\
\bm{\beta}_i&=& \frac{D_L D_{iS}}{D_i D_{LS}}~\bm{\beta}_1~~~~
\left|\left|\frac{\partial\bm{\beta}_i}{\partial\bm{\beta}_1}\right|\right|=
\frac{D_L D_{iS}}{D_i D_{LS}}~\mathsf{I}
~~~~n_\mathrm{L}<i<n_\mathrm{S}
\label{approx0}
\end{eqnarray}
where $\bm{\beta}_i$ is the ray position 
in the i-th plane ($\bm{\beta}_S=0$) and $\mathsf{I}$ is the unity matrix. 
Differentiating our Eq.~\ref{multiplane} and substituting 
$||\partial\bm{\beta}_i/\partial\bm{\beta}_1||$ from Eq.~\ref{approx0},
we get (compare \citealt{McCully16} in different notation): 
\begin{eqnarray}
\mathsf{\Gamma}&\equiv &  \mathsf{I}  - 
\left|\left|\frac{\partial\bm{\beta}_S}{\partial\bm{\beta}_1}\right|\right|
= \left|\left|\begin{array}{cc}
\kappa^\mathrm{ext}+\gamma_1^\mathrm{ext} & \gamma_2^\mathrm{ext}  \\
\gamma_2^\mathrm{ext} & \kappa^\mathrm{ext}-\gamma_1^\mathrm{ext}
\end{array} \right|\right|\\
&=&\sum_{i=1}^{n_L}\frac{D_{iS}}{D_{S}}\mathsf{\Gamma}_i^\prime
+\sum_{i=n_L+1}^{n_S-1}
\frac{D_L}{D_i}\frac{D_{iS}}{D_{LS}}\frac{D_{iS}}{D_{OS}}
\mathsf{\Gamma}_i^\prime
\label{weak_all}
\end{eqnarray}
The weak lensing matrix is calculated in the linear approximation with
strong lens influence on ray paths taken in the zeroth approximation.
In the LOS+ENV case all layers $0<i<n_s$ are included in 
Eq.~\ref{weak_all}. In the LOS approach the strong lens plane is omitted
($\mathsf{\Gamma}_i^\prime=0$ for $i=n_L$) and in the ENV case all other
planes are omitted ($\mathsf{\Gamma}_i^\prime=0$ for $i\ne n_L$).
The results for the convergence are shown in Fig.~\ref{kappa}.
The distributions of the shear are shown in Fig.~\ref{gamma}, where they
are compared to the results of model fitting.

The external convergence distribution in Fig.~\ref{kappa} plotted
with solid line for the LOS case
can be compared with Fig.3 of \citet{Collett13}. Our results give
$\kappa^\mathrm{ext}$ closer to zero. 
While they use fixed source redshift $z_\mathrm{S}=1.4$,  we consider
a range of redshifts ($1.23\le z_S \le 2.62$), which should make our
convergences higher. On the other hand they ignore the influence of
strong lensing, while all our beams contain a strong lens at $0.32\le
z_\mathrm{L} \le 0.83$ which is not directly contributing to the
convergence, but makes contribution of layers farther away less
important (Eq.~\ref{weak_all}). For comparison with Fig.~6 of
\citet{Suy13} we show also conditional probability distribution for the
external convergence weighted by the likelihood of
$\gamma^{ext}=0.089\pm0.006$.  The fraction of cases with such a large
shear value (corresponding to overdense lines of sight) 
is small among our simulated models, so the distributions
have to be plotted with broader bins. The visual comparison of our plot
in LOS+ENV case with their Fig.~6 suggests a rough similarity of the
conditional probability distributions of $\kappa^\mathrm{ext}$.
Our $\kappa^\mathrm{ext}$ plots are also qualitatively similar to 
the results of \citet{smith14} shown in their Fig.~5.

\section{Perturbed quad image configurations}
\label{sec:quad}

We start image finding using approximate methods on the grid. We replace 
the point source by an extended surface brightness profile with the Gaussian 
shape and the characteristic radius of a few pixels. The related surface 
luminosity  in the sky is given by:
\begin{equation}
I^\mathrm{obs}(\bm{\beta}_1)=I^\mathrm{src}(\bm{\beta}_S(\bm{\beta}_1))
\end{equation}
The local maxima of the observed surface luminosity are the positions of the 
images of the source center. We find brightness maxima on the grid and use
them as approximate solutions to the lens equation. The improved positions 
are obtained by iterations.

We consider $16 \times 16$ positions 
of the source with both coordinates in the range $0$ to $3^{\prime\prime}/16$.
(This is an ad hoc choice. The considered source positions cover one 
quadrant of the diamond caustics. The number of quad configurations 
produced by a given lens is roughly proportional to the surface area 
inside the caustics. If one neglects the redshift and magnification 
dependence of the selection process it also gives the relative 
probability of observing quad configuration produced by the lens). 
For each source location we numerically find positions, magnifications, and 
relative time delays of all images, which we call an image configuration.

Configurations perturbed by ENV and/or LOS effects have different 
image positions, relative time delays and magnifications as compared to 
UNI case. We quantify these effects using 
an observer-lens distance estimate based 
on the measurements of relative time delays for two images, their positions, 
and the knowledge of the lens structure. Such methods of distance 
measurements have been employed by \citet{paraficz09} for SIS lenses and by
\citet{jee15} for lenses with power-law density profile. The latter authors 
are mostly interested in the influence of the lens model 
on the distance estimate, while we are interested in perturbations
external to the lens. For this reason we assume the lens model to be known
and use the strong lens from simulations in the following calculations.

In the UNI approach the time-delay between a pair of images $i$ and $j$ 
can be expressed in many equivalent forms, in particular as 
(compare \citealp{jee15}):
\begin{equation}
c\Delta t_{ij}=
(1+z_\mathrm{L})d_\mathrm{OL}
\left(\frac{1}{2}(\bm{\beta}_i-\bm{\beta}_j)(\bm{\alpha}_i+\bm{\alpha}_j)
-\int_{\bm{\beta}_j}^{\bm{\beta}_i} 
\bm{\alpha}(\bm{\beta})\cdot\mathrm{d}\bm{\beta}\right)
\label{addeq}
\end{equation}
where $z_\mathrm{L}$ is the lens redshift, 
$d_\mathrm{OL}$ -- the lens angular diameter distance, 
$\bm{\beta}_i$, $\bm{\beta}_j$ -- the image positions, 
$\bm{\alpha}_i$, $\bm{\alpha}_j$ -- the respective deflection angles.
The formula is valid for any lens model; in our case the integral 
can be expressed analytically using the NSIE potential \citep{K06}.

For a measured relative time delay and image positions one can 
calculate the distance $d_\mathrm{OL}$ from the above equation 
for a lens of known matter distribution in a uniform universe. 
Using the same lens model and applying it to  
the perturbed time delays and image positions 
(resulting from LOS and ENV effects)
one can only estimate the distance, obtaining some  
approximate $d_\mathrm{OL}^\mathrm{est}$. 

\begin{figure}
\includegraphics[width=84mm]{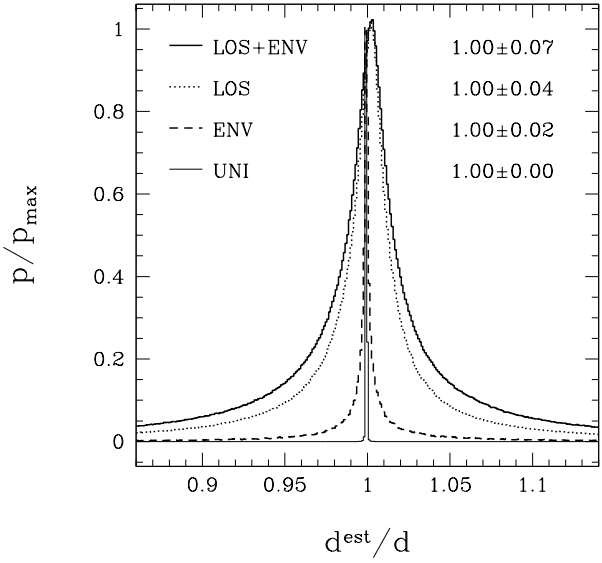}
\caption{Quality of the lens distance estimate. 
The ratios of the estimated to the true values of the lens distance 
are shown as histograms for the LOS+ENV (solid), LOS (dotted), 
ENV (dashed), and UNI (thin solid) cases. The median and half-width of a
central region containing 68\% of the distributions are also given in 
the figure. 
}
\label{add}
\end{figure}

For easier comparison of the perturbed/unperturbed pairs of images 
we use the properties of corresponding deformation matrix $\mathsf{A}$:
\begin{equation}
\mathsf{A}(\bm{\beta})\equiv 
\left|\left|\frac{\partial\bm{\beta}_S}{\partial\bm{\beta}}\right|\right|
\equiv \left|\left|\begin{array}{cc}
1-\kappa-\gamma_1 & -\gamma_2  \\
-\gamma_2         & 1-\kappa+\gamma_1
\end{array} \right|\right|
\label{matrix}
\end{equation}
where $\bm{\beta}_S$ is the source position corresponding to 
the image position $\bm{\beta}$, $\kappa$ is the convergence, 
$\gamma_1$, $\gamma_2$ are the shear components, 
and $\gamma\equiv\sqrt{\gamma_1^2+\gamma_2^2}$ is the shear value.
Using the determinant and trace values 
of matrix $\mathsf{A}$ one can classify images into I -- III categories 
(eg \citealp{K06}):
\begin{eqnarray}
\mathrm{I:} & \det\mathsf{A} >0 & \mathrm{tr}\mathsf{A}>0\\
\mathrm{II:} & \det\mathsf{A} <0 & \\
\mathrm{III:} & \det\mathsf{A} >0 & \mathrm{tr}\mathsf{A}<0
\end{eqnarray}
The quad configurations consist of two type I images (corresponding 
to the minima of the Fermat potential) and two type II images (corresponding 
to the saddle points). The fifth image (if {\it observed}) is of the III type 
(the maximum). Thus a pair of type I (II) perturbed images corresponds 
to and should be compared with the pair of the same type of unperturbed images.
The types of the {\it observed} images may be difficult to assess unless 
they have resolved radio structures. Classification is straightforward when  
using a lens model.

We have found the distributions of the ratios 
$d_\mathrm{OL}^\mathrm{est}/d_\mathrm{OL}$ for configurations perturbed by 
the ENV or/and LOS effects. They are shown in Fig.~\ref{add}. We use
image classification to compare the pairs of the same kind, but we show 
the joint distributions of the distance ratios for type I and type II pairs.
The plots are also labeled using 
the medians and half-widths of regions containing 68\% of the 
distributions shown. 
As can be seen in the figure, the median estimated distances are 
close to their true values. The small scatter in UNI results is of
numerical origin: image finding is done numerically, based on
deflections approximated from the grid.

\section{Simplified fits}
\label{sec:fits}

For each quad image configuration we attempt a fit using our model of an
elliptical lens 
with external shear in a uniform universe model with Hubble constant treated 
as a free parameter. We consider the results with acceptable fits only.
This approach exactly follows that of Paper II, but the sample of lenses
and of image configurations is constructed in a different way.

We use $\chi^2$ statistic to check the quality of fits. The
detailed description of $\chi^2$ as a function of model parameters and
simulated observables is given in Paper II. Each investigated configuration
has 4 image positions 1 galaxy-lens position, 3 relative time delays and
3 image flux ratios - a total of 16 measured quantities. Our models have
4 intrinsic lens parameters ($\alpha_0$, $q$, $r_2$, and the position
angle), 1 galaxy position, 2 components of the shear, 1 source position and
the Hubble constant value as parameters - a total of 11 parameters,
leaving 5 degrees of freedom ($DOF$. We reject models with 
$\chi^2>11.07$ (95\% confidence for $DOF=5$). 
When we fix some of the parameters ($\gamma_1$, $\gamma_2$, $\alpha_0$ -
see below), $DOF$ increases accordingly.

Because of the perturbations by ENV and LOS the fitted models do not reproduce
the parameters of the main lens used in simulations (compare also 
Papers I and II). Even considering an isolated lens in a uniform universe 
model one may find acceptable fits with different lens parameters values 
and non vanishing external shear. Such models may be as good fits to the 
simulated data, as the original model used in the simulations. 

\begin{figure}
\includegraphics[width=84mm]{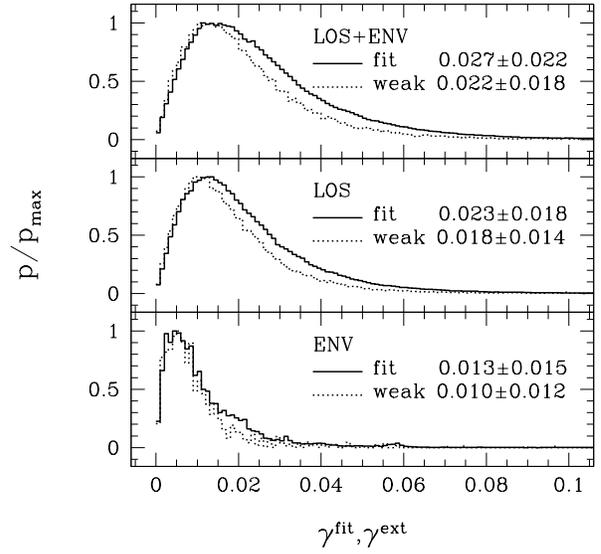}
\caption{Distributions of the external shear. In the upper panel we show
the probability distributions of the fitted $\gamma^\mathrm{fit}$ values for 
acceptable models in the LOS+ENV approach (solid). The distribution of
the weak lensing estimated shear $\gamma^\mathrm{ext}$ for the same set
of models is also shown (dotted). In the middle and lower panels 
the distributions of the fitted (solid) and the estimated (dotted) 
shear are shown for the LOS and ENV approaches. Averaged values and
standard deviations are also shown in the figure.
}
\label{gamma}
\end{figure}

The large number of weak lenses combinations considered 
in the present investigation (as compared to Papers I and II) makes 
the distributions of fitted parameters generally
smoother, and their statistical properties can be better described.
The number of individual quad image configurations, which have acceptable 
fits, is typically a few times $10^4$ for UNI and ENV cases, and
a few times $10^6$ for LOS and LOS+ENV.

Following \citet{McCully16} we assume that the image positions are measured 
with the accuracy of $0.003~\mathrm{arcsec}$, 
the lens position to $0.003~\mathrm{arcsec}$, 
the flux ratios to $0.05~\mathrm{mag}$,
and the time delays to $1~\mathrm{d}$. 
The redshifts of the lens and source are fixed and we do not consider
the influence of their accuracy on the results.

\subsection{External shear}

We start with the comparison of the fitted values of the external 
shear $\gamma^\mathrm{fit}$ with its value estimated in the weak lensing
approximation $\gamma^\mathrm{ext}$ (Eqs.~\ref{weak_one} -- \ref{weak_all}).
In Fig.~\ref{gamma} (upper panel) we show the probability distributions 
of the estimated and fitted shear values in the LOS+ENV approach. (Only
the acceptable fits and corresponding weak maps combinations are
included). In the middle panel the results for the LOS approach are
compared, and in the lower for the ENV. All plots show reasonable 
agreement between the shapes of the distributions, averages and
dispersions. The detailed agreement between the estimated and fitted 
shear values should not be expected as shown by \citet{Wong11}, \citet{KZ04},
and in Paper II.  The apparent deficit of higher shear values in the 
distributions based on weak lensing calculation may result from the
omission of higher order terms; the fits use simulated configurations
based on the full treatment. Simple sum of shear contributions from all
planes gives estimated shear values higher than resulting from the fits.
Removing factor $D_L/D_i$ from the second sum in Eq.~\ref{weak_all}
would give the best agreement between the estimated and the fitted
distributions, but we can see no reason to use such approach. (It
corresponds to the assumption that $\bm{\beta}_i$ decreases linearly
with the comoving distance between the lens and the source, but in fact
it holds for the product $D_i\bm{\beta}_i$.)

\subsection{Lens parameters}

\begin{figure}
\includegraphics[width=84mm]{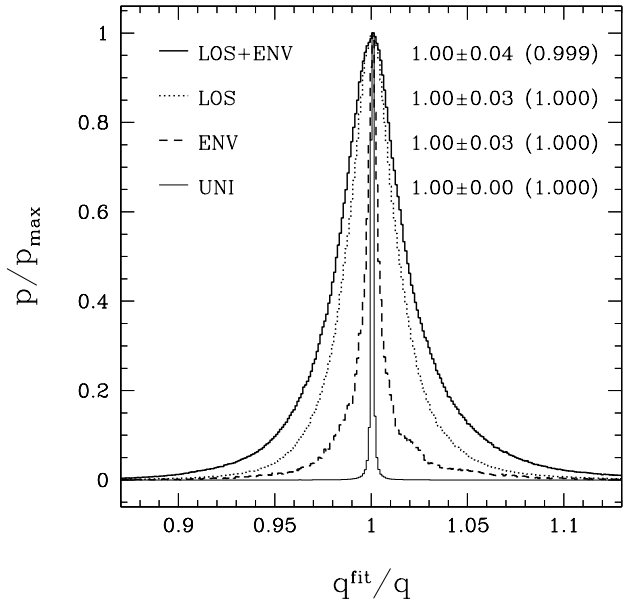}
\caption{Distributions of the fitted lens axis ratio $q^\mathrm{fit}$ 
divided by its original value $q$. The plots related to different 
approaches are labeled in the figure. The average values and their 
standard deviations, with the median values in parentheses, are also shown.
(See the text for details)
}
\label{qqq}
\end{figure}

Similarly we investigate the LOS and ENV effects on values of the fitted
lens axis ratio $q$ and characteristic lens deflection angle $\alpha_0$.
In simulations the images are influenced by shear contributed by LOS, ENV, 
and the elliptical lens itself. 
Similarly all three are sources of convergence. 
Simplified models include only an elliptical lens and an external shear, 
so changing the lens parameters is a natural way to reproduce at least 
some of the effects seen in simulations.

\begin{figure}
\includegraphics[width=84mm]{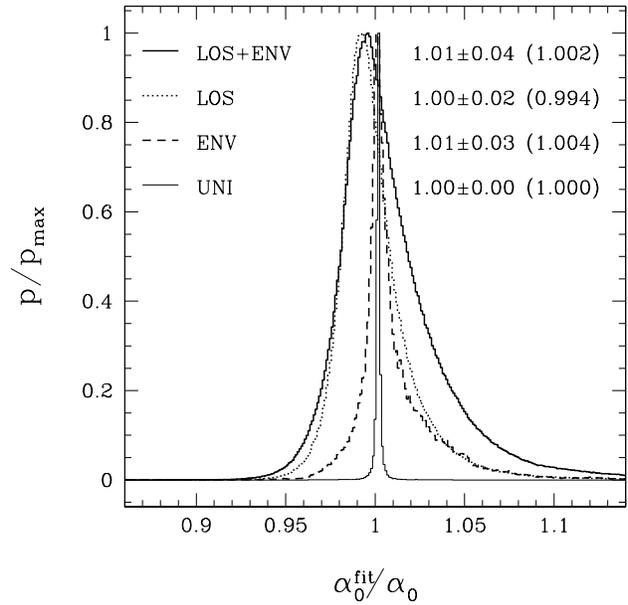}
\caption{Distributions of the ratio of the fitted to original
characteristic deflection angle $\alpha_0^\mathrm{fit}/\alpha_0$. 
The plots related to different 
approaches are labeled in the figures. The average values and their 
standard deviations, with median values in parentheses are also shown.
(See the text for details)
}
\label{alf}
\end{figure}

In Fig.~\ref{qqq} we show the distributions of the ratios of the fitted
to the original parameter values for the lens axis ratio 
($q^\mathrm{fit}/q$). On average the fitted values show no bias
with the dispersion ranging from practically zero (UNI) to $0.03$ 
(ENV, LOS)  to $0.04$ (LOS+ENV). The median values of the distributions
are all very close to one (to better than 0.001). 

In Fig.~\ref{alf} we show the distributions of the ratios of the fitted 
to original characteristic deflection angles 
($\alpha_0^\mathrm{fit}/\alpha_0$). The fitted values are slightly biased 
toward larger values in the ENV and LOS+ENV cases  (by a factor of 1.01)
and unbiased for the UNI and LOS cases. The median values of the
distributions are shifted in the same direction as averages, but remain
closer to the true values.  
The distributions of $\alpha_0^\mathrm{fit}/\alpha_0$  
are not Gaussian and not
symmetric but the standard deviations give a meaningful characteristics 
of their width, similar to the centile analysis.

\subsection{The Hubble constant}

\begin{table*}
 \centering
 \begin{minipage}{168mm}

  \caption{Fitted $H_0$ values}
  \begin{tabular}{@{}ccccc@{}}
  \hline
  Approach 	&  LOS+ENV & LOS & ENV & UNI\\
 \hline
 Full & 73.3$\pm$2.9 (72.7$^{+2.3}_{-1.6}$) 72.7
      & 72.8$\pm$2.2 (72.4$^{+1.6}_{-1.3}$) 72.5
      & 73.5$\pm$2.2 (73.1$^{+1.4}_{-1.7}$) 73.0
      & 73.1$\pm$0.3 (73.0$^{+0.1}_{-0.1}$) 73.0 \\
 $\gamma$,$\kappa$ 
	& 72.9$\pm$1.7 (72.7$^{+1.2}_{-1.2}$) 73.0  
	& 73.0$\pm$1.4 (72.9$^{+0.8}_{-1.0}$) 73.0
	& 73.0$\pm$1.2 (72.8$^{+0.7}_{-0.6}$) 73.0
	& 73.1$\pm$0.3 (73.0$^{+0.1}_{-0.1}$) 73.1  \\
 $\gamma$,$\kappa$,$\alpha_0$
        & 72.9$\pm$1.7 (72.7$^{+1.2}_{-1.0}$) 73.0  
        & 73.0$\pm$1.4 (72.9$^{+0.8}_{-1.1}$) 73.0  
        & 73.0$\pm$1.2 (72.8$^{+0.5}_{-0.4}$) 73.0  
        & 73.1$\pm$0.2 (73.0$^{+0.1}_{-0.1}$) 73.1  \\
 \hline
 \noalign{\vskip3pt}
 \multicolumn{5}{p{16.6cm}}{Note: The statistical characteristics of the 
distributions of the fitted values of the Hubble constant $H_0$ in km/s/Mpc.
The distribution averages plus/minus standard deviations are shown followed 
by the median values in parentheses, and mode value. 
Superscripts and subscripts show the widths of the 
sections containing 34 per cent of the distribution on each side of the median.
The consecutive columns correspond to the 
different descriptions of light propagation 
(LOS: the effects of matter along the line of sight 
included; ENV:the effects of the local lens environment included; LOS+ENV: both;
UNI: none). 
The rows show the results for the different assumptions about lens modeling. 
(Full: no restrictions on the model parameters; $\gamma$, $\kappa$: 
the external shear of the lens assumed equal to the value estimated 
by weak lensing approach, the external convergence from the same estimate 
supplemented to the model;
$\gamma$, $\kappa$, $\alpha_0$: as above plus characteristic deflection angle
$\alpha_0$ assumed equal to the value used in simulations, which 
may correspond to a lens with measured velocity dispersion).
}
\end{tabular}
\end{minipage}
\label{Hubble}
\end{table*}

\begin{figure}
\includegraphics[width=84mm]{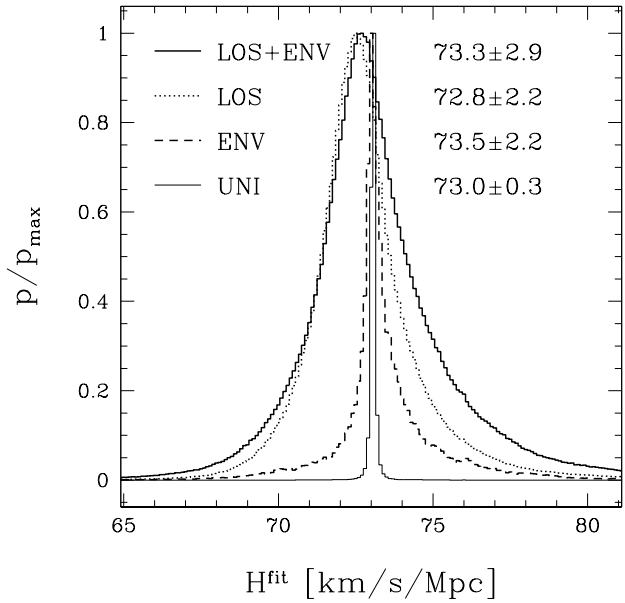}
\caption{Distributions of the fitted Hubble constant values.
The results for the LOS+ENV (solid line) LOS case (dotted)
ENV (dashed) and UNI (thin solid) are shown. The plots show the results
of the full model, with no restrictions on parameters.
Average values and standard deviations are also included in the figure. 
(See Table.~1 for further details and the results for other models.)
}
\label{aahub}
\end{figure}

\begin{figure}
\includegraphics[width=84mm]{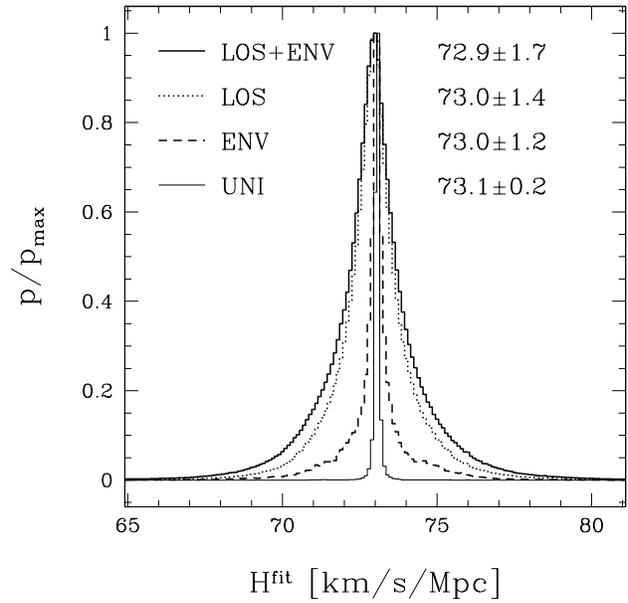}
\caption{Same as on Fig.~\ref{aahub}, but shear parameters of the
models fixed at values estimated for a given strong lens and a combination
of weak lenses, 
$\gamma_1=\gamma_1^\mathrm{ext}$, $\gamma_2=\gamma_2^\mathrm{ext}$, and
the value of the external convergence $\kappa^\mathrm{ext}$ from the
same estimate supplemented. Conventions follow Fig.~\ref{aahub}.
}
\label{bbhub}
\end{figure}

The distributions of fitted $H_0$ values are given in Fig.~\ref{aahub}.
(The {\it true} value used in simulations was $H_0=73~\mathrm{km/s/Mpc}$).
All quad image configurations leading to successful fits are included 
with the same weight, when calculating the distributions, so the 
lenses with larger caustic areas (and larger number of possible 
quad configurations) contribute more. The distributions give the expected 
fitted value of the Hubble constant based on a single observed 
quad lens.

We present the results for all four descriptions of the light propagation. 
The average and median value of the fitted Hubble constant is close 
to the original $H_0$ in all cases (closer than $0.6~\mathrm{km/s/Mpc}$).
The distributions are not Gaussian and not symmetric, each in its own way.
For LOS+ENV and LOS the maximum of the probability distributions are
shifted to lower values

The fitted values of the shear are in good agreement with the
values estimated in the weak lensing approximation (see
Fig.~\ref{gamma}, Eq.~\ref{weak_all}). We have run our lens models
again with fixed shear parameters values
$\gamma_1=\gamma_1^\mathrm{ext}$, $\gamma_2=\gamma_2^\mathrm{ext}$,
and supplementing them with the estimated convergence value
$\kappa^\mathrm{ext}$. The results in Table~1 show, that the
bias in averaged values of the fitted Hubble constant and
their spread are reduced.

We also check the influence of the measurement of the velocity 
dispersion in the lens on our models. 
A known velocity dispersion translates into a known value 
of the deflection angle $\alpha_0$. Fixing this parameter and using 
the estimated values of the shear and convergence we fit our models 
once more. The parameters of calculated $H^\mathrm{fit}$ distributions 
are shown in Table~1 and could be compared with the results of other 
fits.

\begin{figure}
\includegraphics[width=84mm]{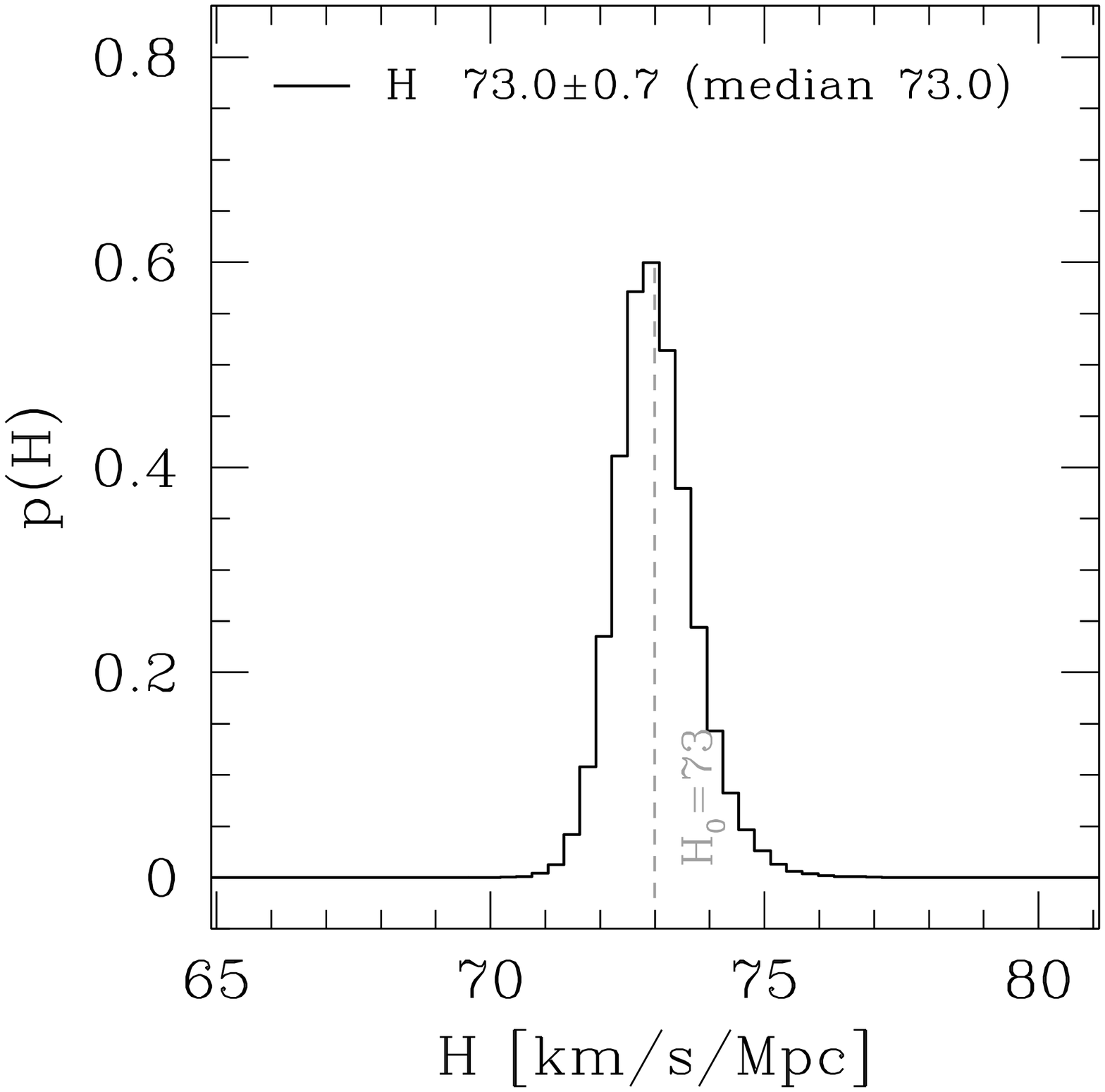}
\caption{Distribution of Hubble constant values $H$ as measured 
  from the samples containing 12 strong lenses each,
  are presented (with outlier rejection employed),
  labeled with the average, dispersion, and median values.
}
\label{hhhub}
\end{figure}

\subsection{Simulated measurement of $H_0$}

Using our data we can simulate the measurement of $H_0$ based on
a random choice of several of our configurations mimicking small 
samples of observed lenses with measured time delays.
The LOS+ENV approach is the most realistic description of light propagation
and we do not show results of other approaches.

We choose from all lenses and all investigated source positions, however we
consider only the cases with quadruple images and acceptable fits. 
When constructing small lens samples we choose a dozen different lenses 
from a total of 1920 used in simulations. The probability of 
a lens to be included  in a sample is proportional to the surface area 
of its caustic, which partially reproduces the probability of a lens 
to produce multiple images.
For a given lens we choose at random one from quad image configurations 
giving satisfactory fits. 

For $N$ chosen configurations which give fitted Hubble constant values
$H^i$ we get the sample average $H$, sample estimated uncertainty
$\sigma$, and the estimated error of the average $\Delta H$: 
\begin{equation}
H=\frac{1}{N}~\sum_{i=1}^N~H^i
~~~~~~\sigma^2=\frac{\sum~(H-H^i)^2}{N-1}
~~~~~~\Delta H=\sqrt{\frac{\sigma^2}{N}}.
\label{least}
\end{equation}

\begin{figure}
\includegraphics[width=84mm]{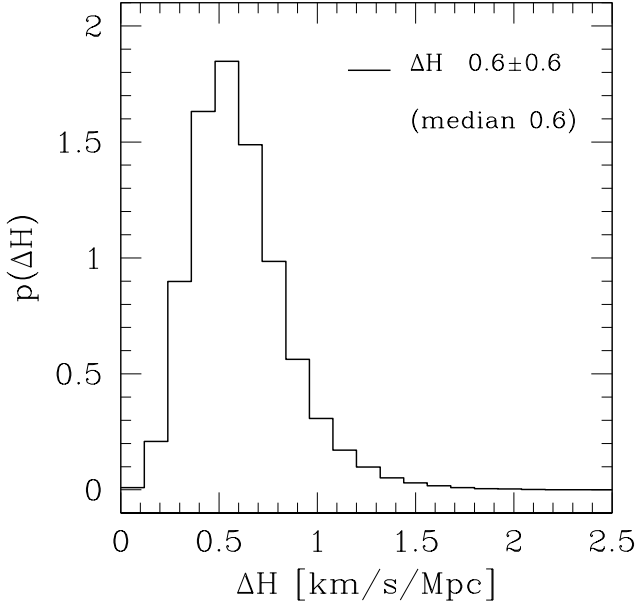}
\caption{Distributions of estimated errors $\Delta H$ of Hubble constant
measurement based on samples of 12 strong lenses with outliers rejection 
are presented.}
\label{sssig}
\end{figure}

We draw a large number of samples of twelve quad lenses each 
and check the results of $H_0$ estimates. 
Note, that the distribution of fitted values of Hubble constant 
($H^{fit}$), as presented on the Fig.~\ref{aahub}, has  wide wings.
The statistics used to form Eq.~\ref{least}
stems from the considerations 
with assumed normal distribution of measurement errors. 
Hence, the measurements coming from the wings of the distribution 
look like outliers and influence the results considerably.
In order to better simulate the real experiment (of measurement 
of $H_0$ from a small sample of real observed gravitational lenses), 
we introduce an outlier-rejection procedure.
For a sample of 12 fiducial measurements, we choose the one that is 
furthest from the mean in the measures of its uncertainty.
Then, we recalculate the mean and uncertainty of the remaining measurements. 
With such rescaled uncertainties, we test if the initially chosen 
measurement is further than 3 of its uncertainties from the mean 
(a $3\sigma$ outlier).
If so, we remove it from the sample and repeat the process until 
no further outliers are found.
In 52\% of random draws (of 12 measurements) no outliers are found, 
in 33\% one measurement is rejected, in 11\% two, and in 3\%  
 of draws three measurements are rejected as outliers.
We calculate the  mean ($H$) of the remaining measurements 
using Eq.~\ref{least}. 
We show the distributions of calculated $H$ in 
Fig.~\ref{hhhub}, and of $\Delta H$ in Fig.~\ref{sssig}.

The calculation shows that 68\% of 
$H$ values lie within $\pm 0.7~\mathrm{km/s/Mpc}$ of $H_0$ 
Thus a twelve lens sample with measured time delays should give consistent 
estimate of the Hubble constant with $\sim$ 1\% relative accuracy 
(``one sigma''). (Of course only the effects of matter outside the main
lens are considered here. The strong lens model is assumed to have a
NSIE profile.)

\subsection{Strong lenses along LOS}

The weak lensing due to LOS  perturbs image configurations. 
Even a weak perturbation, when acting on a configuration with source close
to a caustic, may change the number of images or change image properties
to an extend preventing their successful modeling. The strong perturbation 
may give such effects in any case.

We check the effect in the following way. Any of the weak lensing maps 
from  a given atlas is used many times in different combinations 
with maps from other layers. Using weak lensing maps from one atlas we 
investigate  LOS effects for 160 strong lenses and many source positions.
If for an
isolated lens (or a lens with neighbours) and some given source position 
there exists a quad configuration giving a successful simplified fit, 
we check the existence of quad configurations and their fits, when 
perturbations by LOS are applied. We list all weak maps involved in
failures to get a quad configuration and/or in failures to get a successful 
fit for cases where quads exist. Then we look for maps, which are 
overrepresented among those taking part in failures. In each layer we find 
a map,  which is most frequently involved in changing the number of images 
or in preventing successful fits. We list cases, where the same map in 
a given layer is most responsible for both kinds of failures. 
In each of 12 atlases used in calculations there are 2 -- 4 such maps 
and they 
take part in failures $\sim 3$ times more frequently then average. We then
visually examine the suspected maps, usually finding that the deflection and
time delay patterns are affected by a strong lens close 
to the line of sight.

\section{Discussion and conclusions}
\label{sec:concl}

In this paper we have followed the methods of Papers I and II in simulating
multiple image configurations of strong gravitational lenses based on the 
results of Millennium Simulation (\citealt{Spr05}, \citealt{ls06}).
As compared to our previous papers we have changed the method of main lenses 
selection. We now look for strong haloes in randomly chosen sub regions of 
Millennium Simulation volume instead of looking for strong lenses inside 
a simulated wide, low resolution beam of rays. When finding an appropriate 
candidate we map the matter distribution in its vicinity obtaining a model
of strong lens with its environment. When looking for lenses we do not 
place any extra requirements on their possible environments. (Thus we may
be missing some important property of the observed lenses - compare 
\citealp{Will06}). 

We use the mass density profile of the lens from NSIE model and treat it
as fixed in our calculations. The parameters of the lens may vary but it
remains an isothermal ellipsoid.
This guarantees the analytical formulae for the deflection
angle and gravitational time delay in our models, but also removes the
important source of the uncertainty, the lens mass profile slope
degeneracy (compare \citealp{Suy12}). Thus our estimates of the accuracy
of the Hubble constant measurements do not include the uncertainty of
the mass profile slope. We do not include external convergence as a
parameter in our models.

We have improved our description of matter along the line of sight 
investigating a large number of cases. In the multiplane approach we use 
different combinations of weak deflection maps between the observer 
and the strong  lens and between the lens and the 
source for the fixed strong lensing map and the source position.
Thus the influence of different 
lines of sight on a given source--lens--observer configuration can be 
examined statistically based on ray tracing. We do not attempt to obtain
or employ the relations between the shear, convergence and galaxy 
over-density near the lens as \citet{Suy13}, \citet{Suy10} do for 
individual lenses.

Comparison of the typical shear and convergence values used in our
simulations (Fig.~\ref{kappa}, Fig.~\ref{gamma}) with the values
employed in modeling of real quad lenses (\citealp{Wong11}, \citealp{Suy10},
\citet{Suy13}) suggests that our results may have no direct application to
the extreme cases analyzed there. According to our study the shear and
convergence along random lines of sight are rather small. On the other
hand higher than average density of matter along any line of sight makes
strong lensing more likely to be observed and may serve as a selection
effect not included in our study.

According to our simulations the effects of matter along the line of
sight and in the strong lens vicinity are on average weak and do not
produce a substantial bias in the expected value of the Hubble constant,
which remains close to the true value within $\sim
0.5~\mathrm{km/s/Mpc}$. The uncertainty of a measurements based on a
single lens is $\sigma\approx 3~\mathrm{km/s/Mpc}$, LOS and ENV contributing
roughly $2~\mathrm{km/s/Mpc}$ each. The distribution is not
Gaussian and a probability obtaining a value with $>3~\sigma$ error from
a single lens is $\sim 3$ per cent. Using a sample of dozen lenses gives
the average with $\sim 0.7~\mathrm{km/s/Mpc}$ uncertainty.

The external shear can be found by fitting a model to the image
configuration or by estimating it from the simulation of light
propagation (Eq.~\ref{weak_all}). Since both methods produce similar
results  (Fig.~\ref{gamma}), we make a numerical experiment
substituting the estimated values as model parameters. We also
supplement our model with the estimated value of the convergence.
As a result we get the new distributions of fitted Hubble constant,
which are much more symmetric and narrower as compared to the approach
not restricting model parameters. The average values and medians of the
$H^\mathrm{fit}$ are now at $\le 0.2~\mathrm{km/s/Mpc}$ from the true 
value and the uncertainty $\sim 2~\mathrm{km/s/Mpc}$ - compare Table.~1.
This suggests that on average the uncertainty of the Hubble 
constant measurement resulting from the unknow shear and convergence 
values is $\sim 2~\mathrm{km/s/Mpc}$. It also shows that estimates 
of the shear and convergence values independent of lens modeling may 
substantially improve the accuracy of $H_0$ measurements, a fact well 
known to the strong lensing community.

Results of Sec.~\ref{sec:quad} give a quantitative estimate of the 
LOS and ENV 
effects on the time delay distance measurements. We find that using a pair 
of images and assuming the lens model to be fixed we get $\sim 0.07$ 
1-$\sigma$ relative error in the results (Fig.~2). Distances based on fits 
to the image configurations  and derived Hubble constant values 
(Fig.~\ref{aahub})  have the accuracy of $\sim 4$ per cent. 
Thus our time delay distance estimates remain tentative. 

We have simulated a measurement of the Hubble constant based on 
a small sample of observed time delay lenses drawing the objects from our
huge set of fitted lens models. Each sample includes image configurations 
belonging to different lenses. Drawing the samples large number of times 
and using a method of outliers elimination, we get distributions of 
sample-averaged Hubble constant measurements (Figs.~\ref{hhhub}, 
~\ref{sssig}). Our experiment shows  that in 68\% of cases the result 
lies within $\sim 0.7~\mathrm{km/s/Mpc}$ or $\sim 1$\% from the true 
(used in  the light propagation simulations) value, which should be 
treated as a line of sight and lens environment only contribution 
to the error in $H_0$.

\section*{Acknowledgments}
We are grateful to the Anonymous Referee, whose critical remarks greatly 
improved the paper.
The Millennium Simulation databases used in this paper
and the web application providing on-line access to them were constructed
as part of the activities of the German Astrophysical Virtual Observatory.
We are grateful to Volker Springel for providing us with the smoothed
Millennium density distribution in the early stage of this project. 
This work has been supported in part by the Polish 
National Science Centre grant N N203 581540.


\begin{thebibliography}{99}
\bibitem[\protect\citeauthoryear{Ade et al.}{2014}]{planckXVI}
Ade, P.~A.~R. et al.\ 2014, A\&A, 571, A16
\bibitem[\protect\citeauthoryear{Auger et al.}{2007}]{Aug07}
Auger, M.W., Fassnacht, C.D., Abrahamse, A.L., Lubin, L.M.,
and Squires, G.K., 2007, AJ, 134, 668 
\bibitem[\protect\citeauthoryear{Bar-Kana}{1996}]{barkana96}
Bar-Kana R., 1996, ApJ, 468, 17
\bibitem[\protect\citeauthoryear{Bertone et al.}{2007}]{b4}
Bertone S., De Lucia G., Thomas P.A., 2007, MNRAS, 379, 1143
\bibitem[\protect\citeauthoryear{Carbone et al.}{2008}]{carbone08}
Carbone, C., Springel, V., Baccigalupi, C., Bartelmann, M., 
Matarrese, S., 2008, MNRAS, 388, 1618
\bibitem[\protect\citeauthoryear{Chen et al.}{2003}]{ChKK03}
Chen J., Kravtsov A.V., Keeton C.R., 2003, ApJ, 592, 24
\bibitem[\protect\citeauthoryear{Collett et al.}{2013}]{Collett13}
Collett, T.E. et al., 2013, MNRAS, 432, 679
\bibitem[\protect\citeauthoryear{D'Aloisio \& Natarajan}{2011}]{AN11}
D'Aloisio A., Natarajan P., 2011, MNRAS, 411, 1628
\bibitem[\protect\citeauthoryear{D'Aloisio et al.}{2013}]{dAloisio13}
D'Aloisio A., Natarajan P., Shapiro, P.R., 2014, MNRAS, 445, 3581
\bibitem[\protect\citeauthoryear{De Lucia \& Blaizot}{2007}]{b11}
De Lucia G., Blaizot J., 2007, MNRAS, 375, 2
\bibitem[\protect\citeauthoryear{Fadely et al.}{2010}]{fadely10}
Fadely, R., Keeton, C.R., Nakajima, R., Bernstein, G.M., 2010, ApJ, 711, 246
\bibitem[\protect\citeauthoryear{Freedman \& Madore}{2010}]{freedmad10}
Freedman, W.L., Madore, B.F., 2010, ARAA, 48, 673
\bibitem[\protect\citeauthoryear{Gorenstein et al.}{1988}]{gfs88}
Gorenstein, M.V., Falco, E.E., Shapiro, I.I., 1988, ApJ, 327, 693
\bibitem[\protect\citeauthoryear{Hinshaw et al.}{2013}]{WMAP9}
Hinshaw, G. et al., 2013, ApJS, 208, 19
\bibitem[\protect\citeauthoryear{Hilbert et al.}{2007}]{hilbert07}
Hilbert, S., White, S.D., Hartlap, J., Schneider, P.,2007, MNRAS, 382, 121
\bibitem[\protect\citeauthoryear{Hilbert et al.}{2009}]{hilbert09}
Hilbert, S., White, S.D., Hartlap, J., Schneider, P.,2009, A\&A, 499, 31
\bibitem[\protect\citeauthoryear{Jaroszynski \& Kostrzewa-Rutkowska}
{2012}]{JK12}
Jaroszynski M., Kostrzewa-Rutkowska Z., 2012, MNRAS, 424, 325 (Paper I)
\bibitem[\protect\citeauthoryear{Jaroszynski \& Kostrzewa-Rutkowska}
{2014}]{JK14}
Jaroszynski M., Kostrzewa-Rutkowska Z., 2014, MNRAS, 439, 2432 (Paper II)
\bibitem[\protect\citeauthoryear{Jee et al.}{2015}]{jee15}
Jee, I., Komatsu, E., Suyu, S.H., 2015, JCAP, 11, 033
\bibitem[\protect\citeauthoryear{Keeton et al.}{1997}]{KKS97}
Keeton C. R., Kochanek C. S., Seljak U., 1997, ApJ, 487, 42
\bibitem[\protect\citeauthoryear{Keeton \& Zabludoff}{2004}]{KZ04}
Keeton C. R., Zabludoff, A.I., 2004, ApJ, 612, 660
\bibitem[\protect\citeauthoryear{Kimm \& Yi}{2007}]{KY07}
MKimm T., Yi S.K., 2007, ApJ, 670, 1048
\bibitem[\protect\citeauthoryear{Kochanek}{2006}]{K06}
Kochanek C.S., 2006, in: Gravitational Lensing: Strong, Weak and
Micro, Saas-Fee Advanced Course 33, Springer, Berlin
\bibitem[\protect\citeauthoryear{Kochanek \& Apostolakis}{1988}]{KA88}
Kochanek C.S., Apostolakis J., 1988, MNRAS, 235, 1073
\bibitem[\protect\citeauthoryear{Kormann et al.}{1994}]{KSB94}
Kormann R., Schneider P., Bartelmann M., 1994, A\&A, 284, 285
\bibitem[\protect\citeauthoryear{Lemson \& Springel}{2006}]{ls06}
Lemson G. , Springel V., 2006, Astr.Soc.Pac.Conf., 351, 212
\bibitem[\protect\citeauthoryear{McCully et al.}{2014}]{McCully14}
McCully, C., Keeton C., Wong, K.C., Zabludoff A.,2014, MNRAS, 443, 3631
\bibitem[\protect\citeauthoryear{McCully et al.}{2016}]{McCully16}
McCully, C., Keeton C., Wong, K.C., Zabludoff A.,2016, arXiv:1601.05417
\bibitem[\protect\citeauthoryear{Momcheva et al.}{2006}]{Mom06}
Momcheva, I., Williams K., Keeton C., Zabludoff A.,2006, 641, 169
\bibitem[\protect\citeauthoryear{Oguri}{2007}]{oguri07}
Oguri, M.,2007, ApJ, 660, 1
\bibitem[\protect\citeauthoryear{Paraficz \& Hjorth}{2009}]{paraficz09}
Paraficz, D., Hjorth, J., 2009, A\&A, 507, L49
\bibitem[\protect\citeauthoryear{Paraficz \& Hjorth}{2010}]{paraficz10}
Paraficz, D., Hjorth, J., 2010, ApJ, 712, 1378
\bibitem[\protect\citeauthoryear{Rathna Kumar et al.}{2015}]{rathna15}
Rathna Kumar, S., Stalin, C.S., Prabhu, T.P., 2015, A\&A, 580, A38
\bibitem[\protect\citeauthoryear{Refsdal}{1964}]{refsdal64}
Refsdal, S., 1964, MNRAS, 128, 307
\bibitem[\protect\citeauthoryear{Riess et al.}{2011}]{riess11}
Riess, A.G., 2011, ApJ, 730, 119
\bibitem[\protect\citeauthoryear{Schneider et al.}{1992}]{SEF92}
Schneider P., Ehlers J., Falco E.E., 1992, ``Gravitational Lenses'',
Springer-Verlag Berlin Heidelberg New York
\bibitem[\protect\citeauthoryear{Schneider \& Weiss}{1988}]{SW88} 
Schneider P., Weiss A., 1988, ApJ, 327, 526
\bibitem[\protect\citeauthoryear{Springel et al.}{2005}]{Spr05}
Springel V., White S.D.M., Jenkins A., et al., 2005, Nature, 435, 629
\bibitem[\protect\citeauthoryear{Seitz \& Schneider}{1992}]{SS92}
Seitz S. Schneider P., 1992, A\&A, 265, 1
\bibitem[\protect\citeauthoryear{Schneider}{2014a}]{Sch14a}
Schneider P., 2014, A\&A, 568, L2
\bibitem[\protect\citeauthoryear{Schneider}{2014b}]{Sch14b}
Schneider P., 2014, arXiv:1409.0015
\bibitem[\protect\citeauthoryear{Smith et al.}{2014}]{smith14}
Smith, M., et al., 2014, Apj, 780, 24
\bibitem[\protect\citeauthoryear{Suyu}{2012}]{Suy12}
Suyu S.H., 2012, MNRAS, 426, 868
\bibitem[\protect\citeauthoryear{Suyu et al.}{2013}]{Suy13}
Suyu S.H., Auger M.W., Hilbert, S. et al., 2013, ApJ, 766, 70
\bibitem[\protect\citeauthoryear{Suyu et al.}{2010}]{Suy10}
Suyu S.H., Marshall P.J., Auger M.W., Hilbert, S., Blandford R.D., 
Koopmans L.V.E., Fassnacht C.D., Treu T., 2010, Apj, 711, 201
\bibitem[\protect\citeauthoryear{Wambsganss et al.}{2004}]{WBO04}
Wambsganss J., Bode P., Ostriker J.P., 2004, ApJ, 606, L93
\bibitem[\protect\citeauthoryear{Wambsganss et al.}{2005}]{WBO05}
Wambsganss J., Bode P., Ostriker J.P., 2005, ApJ, 635, L1
\bibitem[\protect\citeauthoryear{Williams et al.}{2006}]{Will06}
Williams K.A., Momcheva I., Keeton C.R., Zabludoff A.I., Lehar J., 
2006, ApJ, 646, 85
\bibitem[\protect\citeauthoryear{Wong et al.}{2011}]{Wong11}
Wong, K.C., Keeton C.R., Williams K.A., Momcheva I.G., Zabludoff A.I.,
2011, ApJ, 726, 84
\end{thebibliography}
\end{document}